\documentclass[aps,pre,twocolumn,floatfix,superscriptaddress]{revtex4-1}
\usepackage{amsmath, amssymb, graphics, mathrsfs, bm, stackrel,bbold}

\usepackage[usenames,dvipsnames]{xcolor}
\definecolor{Highlight}{rgb}{1,1,0.75}
\newcommand{\authormain}{Srividya Iyer-Biswas}
\newcommand{\titlemain}{}
\usepackage{graphicx}
\usepackage{epstopdf}
\usepackage[bookmarks={false}, pdfauthor={\authormain}, pdftitle={\titlemain}]{hyperref}
\hypersetup{colorlinks=true, linkcolor=BrickRed, citecolor=Violet, filecolor=OliveGreen, urlcolor=RoyalBlue, filebordercolor={.8 .8 1}, urlbordercolor={.8 .8 0}}
\usepackage[all]{hypcap}

\usepackage{graphicx}   % Include figure files
\usepackage{dcolumn}    % Align table columns on decimal point
\usepackage{bm}     % bold math
\usepackage[caption=false]{subfig}

\usepackage{dsfont}     % \mathds for identity matrix
\usepackage{color}		% for commenting in color \textcolor

\newcommand\eq[1]{Eq.~(\ref{eq:#1})}

\usepackage[normalem]{ulem}% add \sout strickout

\usepackage{soul}

\newcommand\ba{\begin{array}}
\newcommand\ea{\end{array}}
\newcommand\nn{\nonumber}

\newcommand{\feyn}[1]{#1\kern-0.45em/}

\newcommand{\tto}{\rightarrow}

\renewcommand\l{\lambda}%non-ascii letter
\renewcommand\t{\tau}%accent char

\renewcommand\th{\theta}%latin char

\newcommand\z{\zeta}

\newcommand\la{\langle}
\newcommand\ra{\rangle}
\newcommand\pd{\partial}

\begin{document}
\title{Biological timekeeping in the presence of stochasticity}
\author{Farshid Jafarpour}
\affiliation{Department of Physics and Astronomy, Purdue University, West Lafayette, IN 47907}
\author{Michael Vennettilli}
\affiliation{Department of Physics and Astronomy, Purdue University, West Lafayette, IN 47907}
\author{Srividya Iyer-Biswas}
\email{iyerbiswas@purdue.edu}
\affiliation{Department of Physics and Astronomy, Purdue University, West Lafayette, IN 47907}

%\date{\today}
\begin{abstract}
Causal ordering of key events in the cell cycle is essential for proper functioning of an organism. Yet, it remains a mystery how a specific temporal program of events is maintained despite ineluctable stochasticity in the biochemical dynamics which dictate timing of cellular events. We propose that if a change of cell fate is triggered by the {\em time-integral} of the underlying stochastic biochemical signal, rather than the original signal, then a dramatic improvement in temporal specificity results. Exact analytical results for stochastic models of hourglass-timers and pendulum-clocks, two important paradigms for biological timekeeping, elucidate how temporal specificity is achieved through time-integration. En route,  we introduce a natural representation for time-integrals of stochastic processes, provide an analytical prescription for evaluating corresponding first-passage-time distributions, and uncover a mechanism by which a population of identical cells can spontaneously bifurcate into subpopulations of early and late responders, depending on hierarchy of timescales in the dynamics. Moreover, our approach reveals  how time-integration of stochastic signals may be realized biochemically, through a simple chemical reaction scheme.

 \end{abstract}
%\pacs{\sib{...}} 
\maketitle

\begin{figure}[b!]
\begin{center}
\resizebox{8cm}{!}{\includegraphics{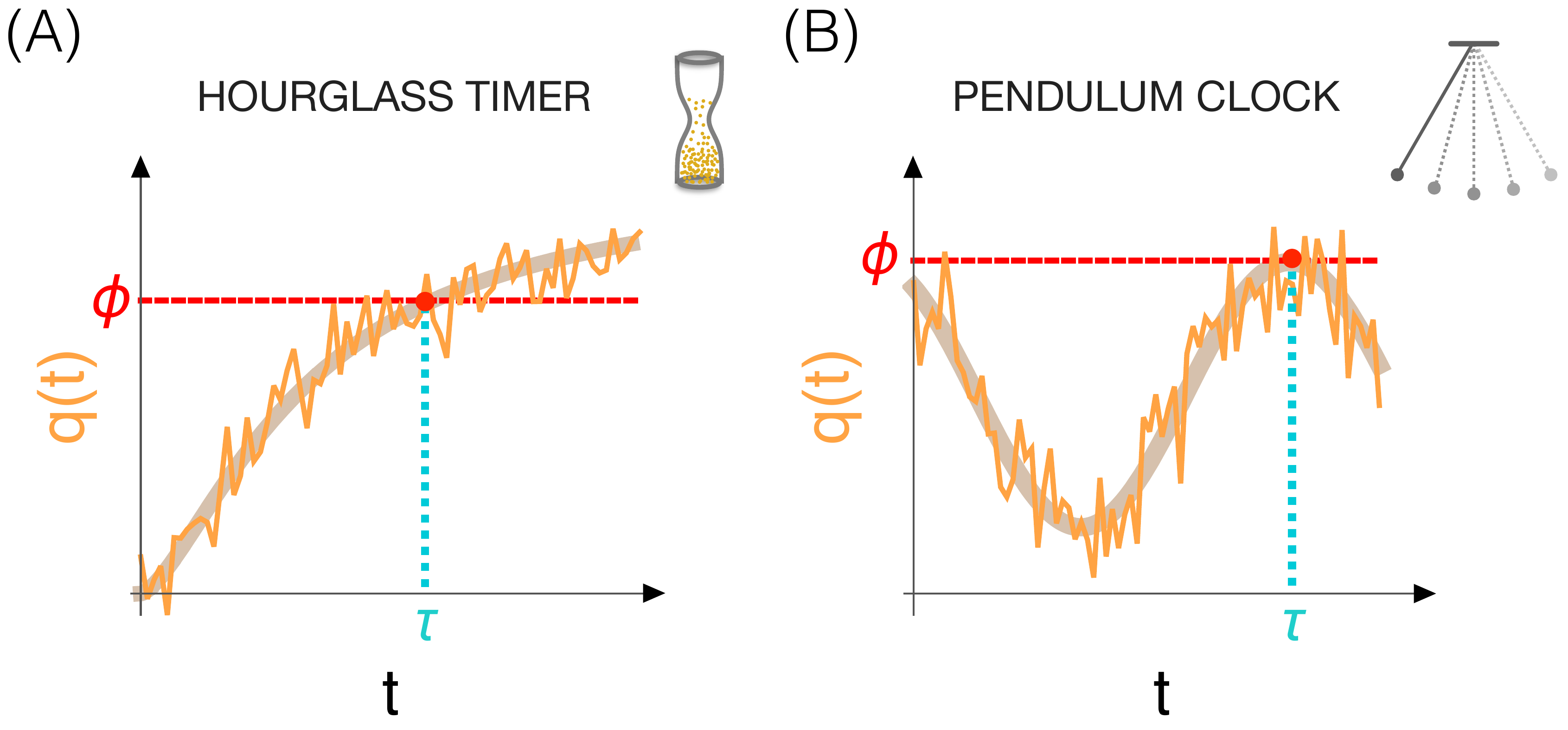}}%inclgphcs[trim=lcm bcm rcm tcm, clip=true, angle=-90]
\caption{{\bf Paradigms of biological timekeeping. } (A) {\bf The hourglass timer}: a cellular dynamical variable representing the biochemical timekeeper, $q(t)$, increases (or decreases) monotonically (beige curve); when it reaches a critical threshold value, $\phi$, at time, $\t$, it triggers the desired cell-fate change. (B) {\bf The pendulum clock}: the biochemical timekeeper, $q(t)$, oscillates periodically, and rhythmically triggers the event of interest at a specific phase of the oscillation, corresponding to a threshold level, $q(t) = \phi$,  after a time interval, $\t$, following the previous event. The orange curves show specific stochastic realizations of $q(t)$ (see accompanying text).}
\label{fig-paradigms}
\end{center}
\end{figure}

Biological clocks are ubiquitous in nature. They govern temporal aspects of biological rhythms and irreversible cell-fate changes~\cite{2001-rensing-ff, 2017-johnson-zv, 1978-cloudsley-thompson-ya, 2017-kumar-ri}. Well known examples include circadian rhythms and cell-division. Typically, these clocks are regulated by biochemicals, whose copy number dynamics dictate when the corresponding biological events occur~\cite{2001-rensing-ff, 2017-johnson-zv, 1978-cloudsley-thompson-ya, 2017-kumar-ri, 2016-iyer-biswas-pz, 2017-pattanayak-ta}.  Characteristic  timescales of biological clocks vary over an extraordinarily broad dynamic range (from order of seconds to hundred years)~\cite{2001-rensing-ff, 2017-johnson-zv, 1978-cloudsley-thompson-ya, 2017-kumar-ri}. However, there are unifying themes, which transcend system-specific details,  in mechanistic aspects of how these clocks function. Using them, chronobiologists have identified two important mechanistic schemes for biological timekeeping~\cite{2001-rensing-ff, 2017-johnson-zv}: {\em (1) the hourglass timer} (see Fig.~\ref{fig-paradigms}A) and {\em (2) the pendulum clock} (see Fig.~\ref{fig-paradigms}B).

It is well appreciated that key biochemical processes are inherently stochastic, causing significant cell to cell variability in their copy numbers, even in a population of isogenic, identically prepared cells. Their effects on molecular, organismal and population level dynamics have been explored in detail, both experimentally and theoretically~\cite{2002-elowitz-vh, 2017-raj-zj,  2005-pedraza-nk, 2009-iyer-biswas-yj, 2009-iyer-biswas-rz, 2014-iyer-biswas-jy, 2012-munsky-jv, 2014-iyer-biswas-lt,  2016-iyer-biswas-tn, 2017-pirjol-fd}. However, corresponding stochasticity in the {\em timing} of key cellular events, whose statistics are governed by fluctuating copy numbers, has not received comparable attention~\cite{2016-iyer-biswas-pz}. In part, this is due to experimental challenges in obtaining high quality time-series data which are amenable to analysis for timing noise: this requires {\em in vivo} measurements at the individual-cell level. Increasingly, this challenge is being overcome through rapid development of live single-cell imaging technologies~\cite{2013-norman-aw, 2014-iyer-biswas-vo, 2015-lambert-sb, 2016-potvin-trottier-sv, 2014-melendez-so, 2014-bisaria-du}.

%%%%%%%%%%%%%%%%%%%%%begin%%%%%%%%%%%%%%%%%%%%
%%%%%%%%%%%%%%%%%%%%%%end%%%%%%%%%%%%%%%%%%%%

An outstanding question in the context of stochastic biological timekeeping is how specific time ordering of key cellular events is achieved, given that underlying biochemical processes rely on noisy regulators with fluctuating outputs. For example, DNA replication inevitably precedes cell division despite cell to cell variability in key determinants of both processes~\cite{2011-shapiro-yz}. Important changes in cell fates are often triggered by thresholded events, i.e., upon the attainment of a critical value of a relevant cellular dynamical variable (e.g., copy number of a protein)~~\cite{2001-rensing-ff, 2017-johnson-zv, 2017-kumar-ri}. Since the governing variable itself fluctuates, there is corresponding variability in the times when the same change occurs in each cell of a population. Thus there is a distribution of  ``first passage'' times (FPT),  namely,  times when the stochastic variable {\em first passes} the threshold value in different cells or ensemble members~\cite{2001-redner-iz, 2016-iyer-biswas-pz}. The question is how temporal specificity, i.e., a tight distribution of these crossing times, is achieved despite underlying stochasticity in governing biochemical processes.

\begin{figure}[t!]
\begin{center}
\resizebox{8cm}{!}{\includegraphics{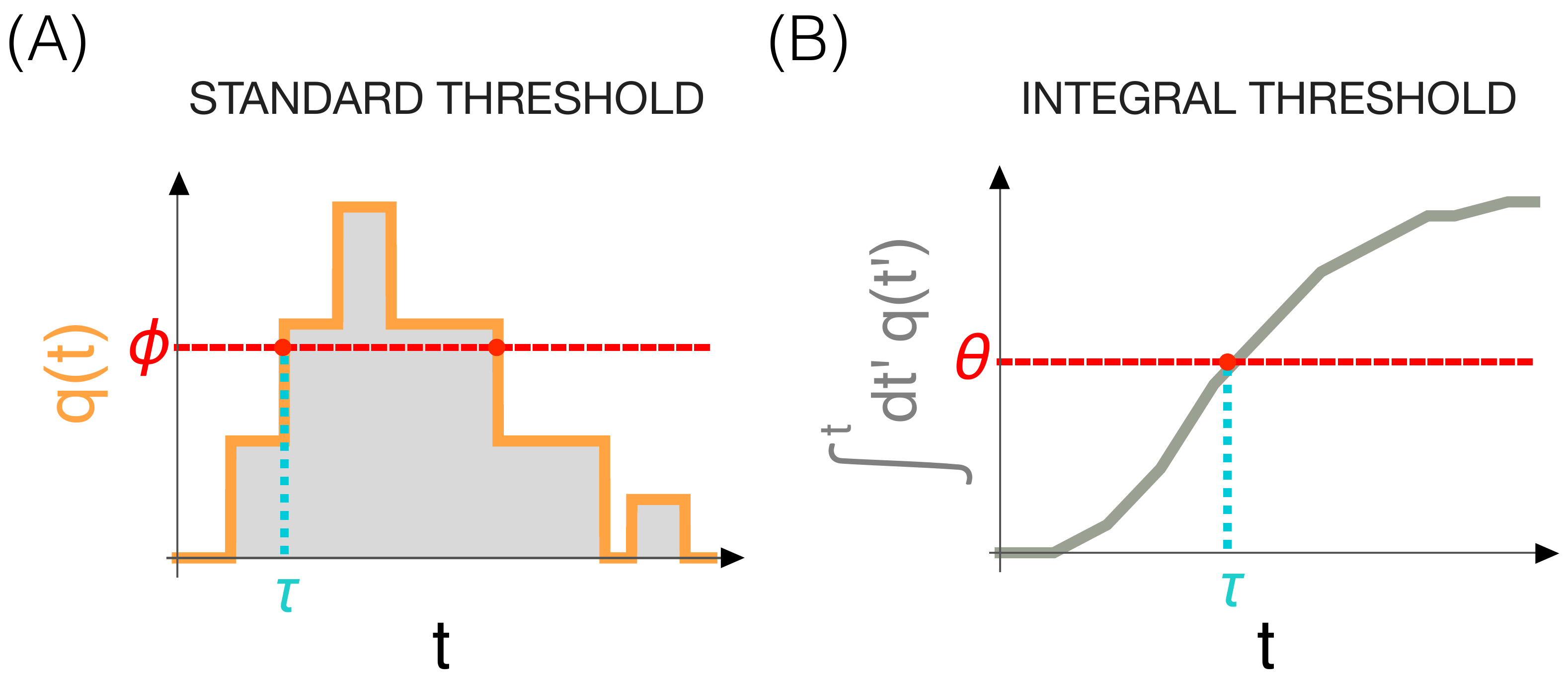}}%inclgphcs[trim=lcm bcm rcm tcm, clip=true, angle=-90]
\caption{{\bf Contrasting integral thresholding with the standard scheme.} (A) A stochastic realization of the copy numbers of the biochemical timekeeper, q(t) (orange curve), has undesired multiple crossing events and missed crossing events (not shown), resulting in large variability in times when the threshold is first crossed in different realizations. In contrast, the time integral of $q(t)$, the gray curve in (B), is monotonically increasing and crosses the corresponding threshold, $\th$, precisely once. Evidently, fluctuations $q(t)$ are suppressed in $\int^{t} dt' q(t')$, thus the integral thresholding scheme significantly reduces variability in threshold crossing times, $\t$, and dramatically improves temporal specificity. (See accompanying text.)}
\label{fig-integral}
\end{center}
\end{figure}

\begin{figure}[t!]
	\centering
	\includegraphics[width=.48\textwidth]{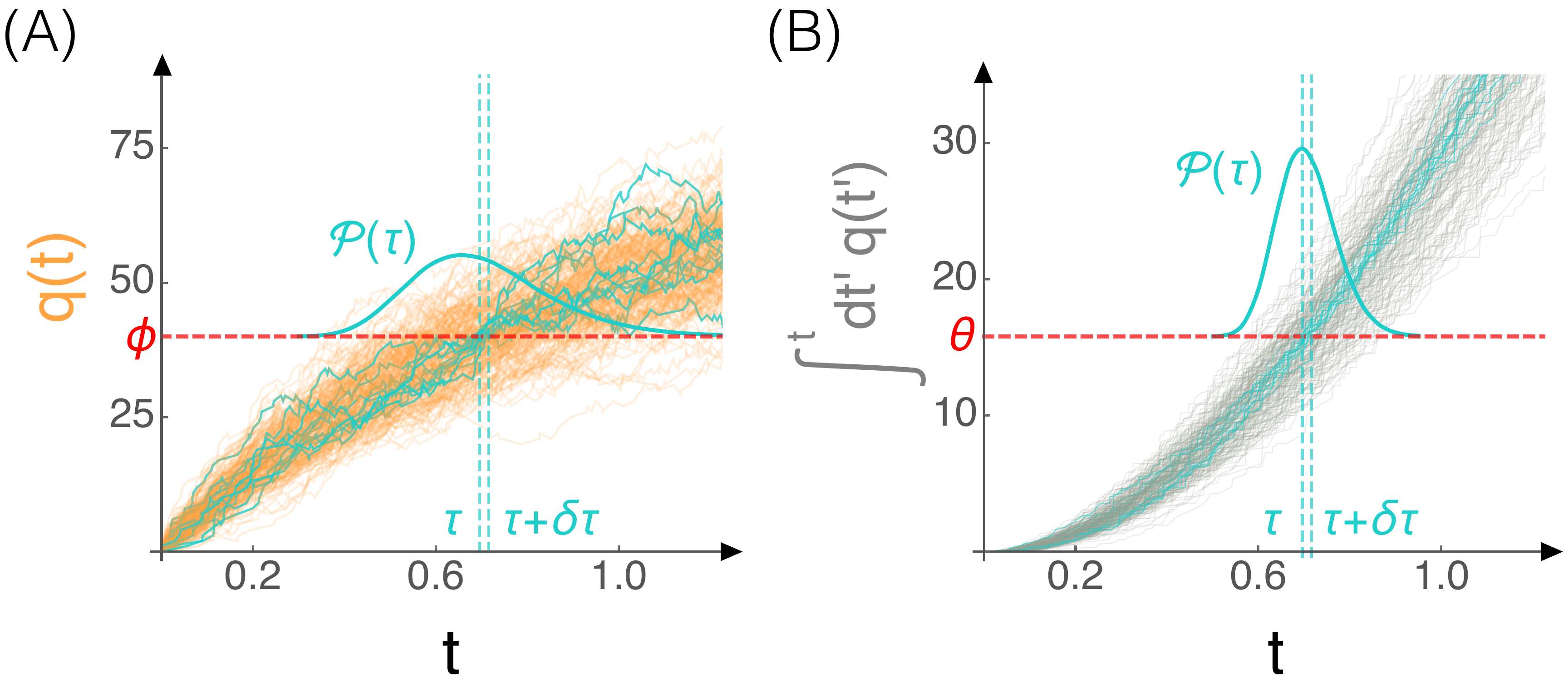}
	\caption{{\bf Applications to stochastic hourglass timers.} (A) Stochastic trajectories, $q(t)$, of a biochemical timekeeper (orange curves) governed by a simple birth-death process, a prototype of a stochastic hourglass timer. Its FPT probability density (bold teal curve), for crossing a standard threshold, $\phi$, is computed by counting the fraction of trajectories crossing the threshold for the first time between times $\t$ and $\t+\delta \t$ (teal trajectories). (B) Corresponding trajectories (gray) of the stochastic time-integral, $\lambda(t) = \int^{t} dt' q(t')$, and the integral's FPT distribution  (bold teal curve). The integral threshold, $\theta$, is chosen to yield the same mean-value, $\la \t \ra$, as in (A). Evidently, thresholding the time-integral results in a significantly narrower FPT distribution, thus improving temporal specificity (see accompanying text for quantification). All timescales are measured in units of $1/k_{d}$; $k_b = 80$,  $\phi=40$, and $\theta = 15.8$.}
	\label{fig:trajectories}
\end{figure}

\begin{figure}[t!]
\begin{center}
\resizebox{8cm}{!}{\includegraphics{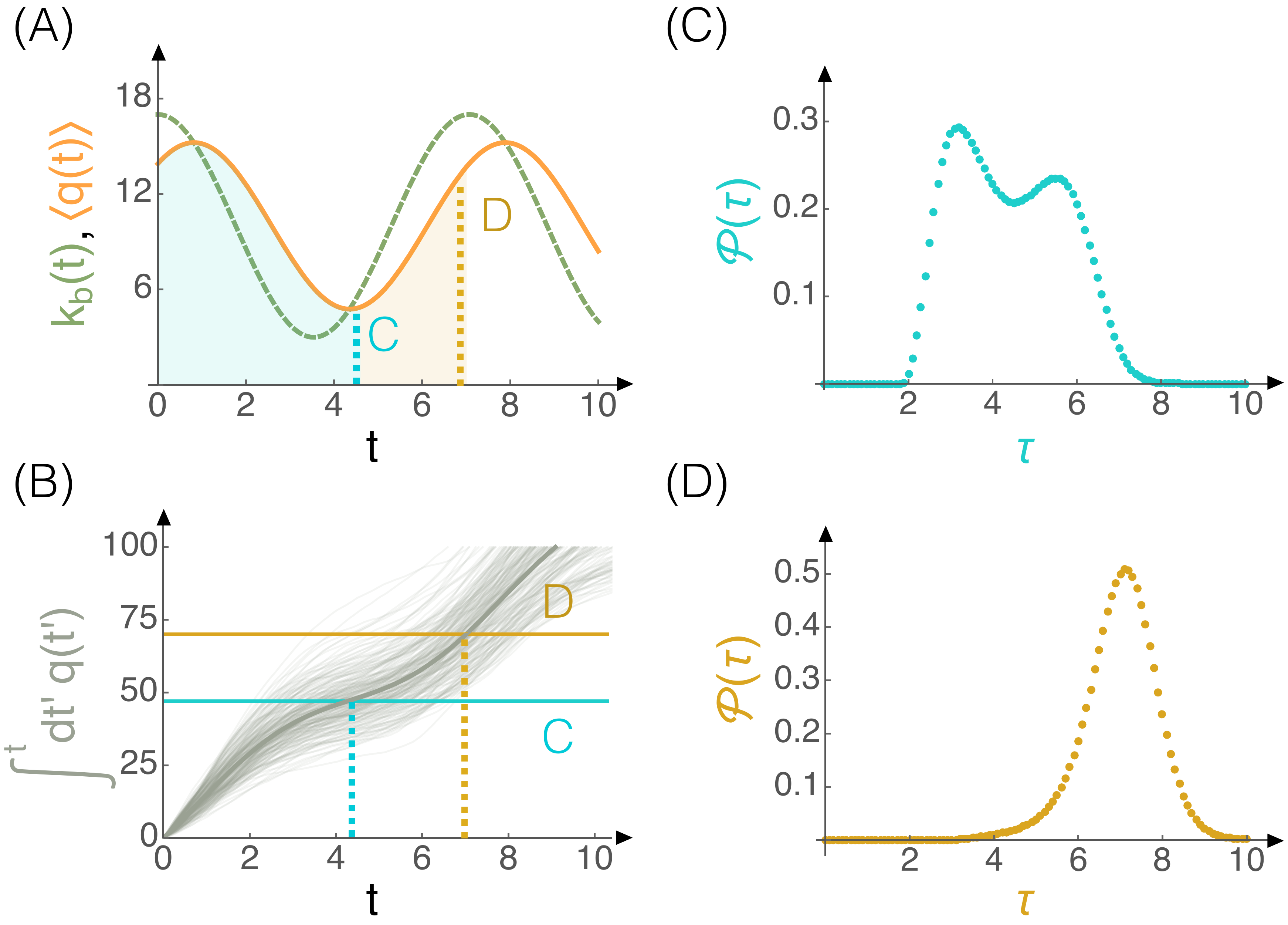}}%inclgphcs[trim=lcm bcm rcm tcm, clip=true, angle=-90]
\caption{{\bf Applications to stochastic pendulum clocks.} (A) The biochemical timekeeper undergoes oscillatory dynamics on average ($\la q(t) \ra$, shown in orange). Its stochastic dynamics are modeled by a simple birth-death process with time-dependent birth rate, $k_{b}(t)$ (dashed green curve).
(B) Stochastic realizations of the time integral, $\int^{t} dt' q(t')$ (gray curves), and their ensemble mean (bold gray curve) are shown. The teal and yellow lines delineate two different integral threshold values; the former corresponds to thresholding at the minimum of the oscillation in $\la q(t) \ra$ while the latter corresponds to a full phase of oscillation (see dotted lines in (A) and (B)). (C) The FPT probability density (teal dots), for the integral to crossing a  threshold value corresponding to the teal line in (B), and the teal area under the curve in (A); a spontaneous bifurcation of the population into early and late responders occurs due a bottleneck effect caused by small number fluctuations (see accompanying text). (D) The FPT probability density (yellow dots), for the integral to crossing a  threshold value corresponding to the yellow line in (B), and the yellow area under the curve in (A); a narrow unimodal distribution is obtained. All timescales are measured in units of $1/k_{d}$; $k_b = 10$,  $\omega=8/9$, $\epsilon = 7/10$ and $\theta = 47 \mbox{ (teal) or }70 \mbox{ (yellow)}$.}
\label{fig-osc}
\end{center}
\end{figure}

In this work we address this issue and show that if the {\em time-integral} of a stochastic biochemical signal is thresholded, rather than the original signal, then a dramatic improvement in temporal specificity results, provided timescales are appropriately chosen (Figs.~\ref{fig-integral}, \ref{fig:trajectories} and \ref{fig-osc}D). By motivating a natural representation for the time-integral of a stochastic process, we provide an analytical prescription for computing FPT distributions of time-integrated signals. We apply this framework to paradigmatic models of biological timekeeping, hourglass timers (Fig.~\ref{fig:trajectories}) and oscillator clocks (Fig.~\ref{fig-osc}), and validate our premise. Further, we show how time-integration of stochastic signals can be realized biochemically, through implementation of a simple chemical reaction scheme. Moreover, these results reveal a mechanism by which a population of identical cells can spontaneously bifurcate into subpopulations of early and late responders, depending on hierarchy of timescales in the dynamics (Fig.~\ref{fig-osc}). We use these results to argue that biochemical time-integration is an attractive prescription for achieving temporal specificity in biological systems.

%predicated

{\bf {\em Biological timekeeping via integral thresholding.}}
We denote the stochastic variable representing the biological timekeeper  by $Q(t)$, its realization by $q(t)$, its time-dependent distribution by $P(q, t)$, the threshold-value by $\th$, the time of the first threshold crossing time by $\t$, and its distribution, i.e., the first-passage-time (FPT) distribution, by $\mathcal{P}(\t)$ (Fig.~\ref{fig-integral}). The mean value, $\la Q(t)\ra$, is an oscillating function of time for pendulum-clock models, and a monotonically increasing (or decreasing) function for hourglass-timer models (Fig.~\ref{fig-paradigms}). The choice of stochastic model for $Q(t)$ will specify statistics of fluctuations around the ensemble-averaged mean-value. Thus the issue is that despite stochasticity in $Q(t)$, one must have a narrow FPT distribution, $\mathcal{P}(\t)$, to ensure  temporal specificity.

%%%%%%%%%%%%%%%%%%%%%begin%%%%%%%%%%%%%%%%%%%%
%%%%%%%%%%%%%%%%%%%%%%end%%%%%%%%%%%%%%%%%%%%

We propose that if the stochastic time-integral, $\int^{t} dt' Q(t')$, is thresholded, instead of  $Q(t)$, then false-positives (undesired crossing events) and false-negatives (missed crossing events) will be naturally eliminated since the time-integral is always monotonically increasing function of time for any stochastic model of $Q(t)$ dynamics. (See Fig.~\ref{fig-integral}.) One will still have cell to cell variability in threshold crossing times, since $Q$'s dynamics are stochastic, nevertheless temporal specificity will be dramatically improved. To test our premise, we introduce a framework to analytically compute the statistics of $\int^{t} dt' Q(t')$, for any general stochastic model of $Q(t)$ dynamics, and also provide a prescription for computing the first-passage-time statistics of the time-integral.

{\bf {\em A natural representation for the time-integral of a stochastic process.}} Evaluation of the statistics of the time-integral of a general stochastic process, $Q(t)$, is a challenging proposition. However, we have found a representation for the time-integral which provides a straightforward and intuitive route to exactly evaluating its time-dependent statistics. 

To physically motivate the representation, we introduce the ``fictitious'' stochastic variable, $R(t)$, produced through the birth process $Q \tto R + Q$, with propensity $k_{r}\, q(t)$, where $k_{r}$ is a rate constant; $P(r, t)$ is the probability of observing $r$ R's at time $t$. Thus the general stochastic model (making no assumptions about the stochastic model governing $Q$ dynamics) consists of:
%%%%%%%%%%%%%%%%%%%%%begin%%%%%%%%%%%%%%%%%%%%
\begin{align}\label{eq:birth}
\begin{split}
&\boxed{\mbox{\em System-specific stochastic model for Q dynamics.}} \\
&\,\,\,\,\,\,\,\,\,\,\,\,\,\,\,\,\,\,\,\,\,\,\,\,\,\,\,\,\,\,\,\,\,\,\,\,\,\,\,\,\,\,\,\,\,\,\,\,Q {\xrightarrow{\;k_r\;}} Q + R.
\end{split}
\end{align}
%%%%%%%%%%%%%%%%%%%%%%end%%%%%%%%%%%%%%%%%%%%

We now relate the statistics of the $R$ variable to the statistics of the time-integral,  $\int^{t} dt' Q(t')$. First, consider a simple limit of the problem: when $q(t)$ is a (deterministic) constant in time, say $q_{0}$, $R$ undergoes a simple birth process, and its distribution is Poisson with mean value, $\l(t) = k_{r} q_{0} t$. Next, consider a slight generalization: when $q(t)$ is a deterministic function of time, $P(r, t)$ is still Poisson distributed~\cite{2009-iyer-biswas-rz, 2009-iyer-biswas-yj}. However, its mean-value, $\l(t)$, becomes a {\em functional} of the deterministic function, $q(t)$: $\l[q(t)] = k_{r} \int^{t} dt' q(t')$. 

For the fully stochastic case, when $Q$'s dynamics are governed by a stochastic model, using the previous result, the distribution $P(r, t)$ must be a superposition of Poisson distributions, since the ensemble can be broken into subpopulations, which share the same stochastic time-course, $q(t)$, and have a corresponding Poisson distribution of $R$. This distribution can therefore be represented as superposition of Poisson distributions with a weighting probability density, $\rho(\l, t)$, which accounts for the frequency with which different trajectories, $q(t)$, arise in the stochastic model~\cite{2009-iyer-biswas-rz, 2009-iyer-biswas-yj, 2014-iyer-biswas-jy}.  Thus, $P(r, t) = \int d\l \,\rho(\l, t) \, {e^{-\l} \l^{r}}/{r!}$. $P(r, t)$ and $\rho(\l, t)$ uniquely determine each other, and this relation can be inverted to find $\rho(\l, t)$, given $P(r, t)$ (which can be computed by solving the master equation for a given model for the dynamics of Q).  However,  setting $k_{r} = 1$, the Poisson parameter, $\l(t)$, is equal to the time-integral, $\int^{t} dt' q(t')$. Thus, $\rho(\l, t)$ is the time-dependent distribution of the time-integral of $Q$! Therefore, the Poisson representation is the natural representation for the time-integral of a stochastic process. It also provides an  analytical route for computing its statistics. 

%%%%%%%%%%%%%%%%%%%%%%%end%%%%%%%%%%%%%%%%%%%%%%
%%%%%%%%%%%%%%%%%%%%%%%end%%%%%%%%%%%%%%%%%%%%%%

We note that for specific stochastic models, we can directly write down a generalized Master Equation for the joint probability of $q$ and the density of its time-integral, $\lambda$. To elucidate this, we consider a class of stochastic models in which $Q$ undergoes birth-death dynamics with a time-dependent birth rate:
\begin{equation}\label{eq:birth_death}
\begin{split}
	\varnothing &\xrightarrow{\;k_b(t)\;} Q\\
	Q &\xrightarrow{\;\;k_d \;\;\;} \varnothing
\end{split}
\end{equation}
Given a system of interest, the functional form of $k_b(t)$ can be chosen to be consistent with the characteristic dynamics of the biochemical timekeeper (see examples below). The joint probability distribution of the number $q$ of $Q$ and its time integral, $\l(t) = \int_0^t q(t') dt'$, is governed by the Master equation:
\begin{align}
\label{eq:master}
	&\pd_{t}{P(q, \l, t)} = k_b(t)[P(q-1, \l, t) - P(q, \l, t)] \nn \\
	&+ k_d[(q+1)P(q+1, \l, t) -  q P(q, \l, t)]  -q \pd_{\l}{P(q, \l, t)}. \nn
\end{align}
The mixed generating function, $G(z, \z, t) \equiv \sum_{q = 0}^\infty \int_0^\infty d\l\, e^{-\z\, \l}\, z^q P(q, \l, t)$, satisfies an analogous Master equation, using which we determine $G(1, -\z, t)$, the moment generating function of $\rho(\l,t)$, and $G(z, 0, t)$, the probability generating function of $P(q, t)$~\cite{2007-van-kampen-pb}. Once $G$ is found, one thus obtains an expression for $\rho(\l,t)$, the probability density of $\l$.

{\bf {\em First-passage-time (FPT) statistics of the time-integral.}}
The FPT distribution, $\mathcal{P}(\t; \th)$, is the distribution of times, $\t$, when the time-integral, $\l$, first passes the threshold value, $\th$.  Irrespective of the details of the stochastic model governing the dynamics of $Q$, its time-integral is a {\em monotonically} increasing function of time, and therefore crosses the threshold, $\theta$, exactly once (see Fig.~\ref{fig-integral}). Thus, we can simply relate the statistics of the time-integral to its FPT statistics~\cite{2014-iyer-biswas-lt, 2016-iyer-biswas-pz}: $\mathcal{P}(\t; \th) = - \pd_{\t} \int_{0}^{\th}\, d\l\, \rho(\l, \t)$. Given a stochastic model for $Q$ dynamics, one can compute the statistics of the time-integral, and its FPT time distribution. 

For the class of models given by Eq.~\ref{eq:birth_death}, we find that the FPT distribution for the stochastic time-integral, $\int^{t} dt' q(t')$, to cross a threshold value, $\th$, is given by an inverse Laplace transform of its moment generating function: 
%%%%%%%%%%%%%%%%%%%%%%begin%%%%%%%%%%%%%%%%%%%%%%
\begin{align}
%\label{eq:FPTD}
%  &\sib{G(1, -\z, \t) \equiv \int_0^\infty d\l\, e^{\z\, \l}\, \rho(\l, \t)}
%	\nn\\
	& \mathcal L\{\mathcal P(\tau, \theta)\} \equiv  \int_{0}^{\infty} d \th \, \mathcal P(\tau, \theta) e^{-\th \z}  \nn \\
		&= -\frac{k_d}{\z} \pd_{\t} \bigg[ F(\z, \t)
		 e^{\frac{\z}{(\z+k_d)}\int_0^\t ds \,k_b(s)  \left(e^{(\z+k_d) (s-\t)}-1\right)} \bigg].
\end{align}
%%%%%%%%%%%%%%%%%%%%%%%end%%%%%%%%%%%%%%%%%%%%%%
$F(\z, t)$ is determined from initial conditions. For instance, when $P(q, \l, 0) = \delta_{q,0}\,\delta(\l)$, $F(\z, t)=1$.\\

%\sib{As a prototype of a mechanistic model of a stochastic hourglass-timer, we consider the model in \eq{birth_death} with the birth rate $k_b$ pick to be constant in time. 
%
%Using exact analytical results, we find that thresholding the time integral rather than the original process significantly narrows the first passage time (FPT) distribution. This distribution is  unimodal, sharply peaked around its mean, with a coefficient of variation that ... with the threshold value.}

%%%%%%%%%%%%%%%%%%%%%%
{\bf {\em Applications to hourglass-timers.}}
As a prototype of a mechanistic model of a stochastic hourglass-timer, we model the dynamics of the biochemical timekeeper, $Q$, with simple birth-death dynamics, i.e., as a simple case of \eq{birth_death} with constant birth rate, $k_b$.  Thus the mean value increases monotonically with time from $0$ to the steady-state value, $q^{*}$: $\la q(t) \ra  = q^{*}[1 - \exp(-k_d t)]$, with $q^{*} = k_{b}/k_{d}$. 

As illustrated in Fig.~\ref{fig:trajectories}, since the original variable and its time-integral are both stochastic,  there is substantial variability in the times, $\t$, when their corresponding threshold values ($\phi$ and $\th$ respectively) are first crossed in different realizations. The variability is characterized by the relative width of the FPT distribution, $\mathcal{P(\t)}$, for each thresholding scheme. For apples to apples comparisons, we constrain the mean first passage time, $\la \t \ra$, to be the same for both cases. 

For improved temporal specificity, $\mathcal{P(\t)}$ should be much narrower for the integral threshold than the standard threshold. As evident in Fig.~\ref{fig:trajectories}, this is true, since the coefficient of variation, i.e., the ratio of standard deviation to mean, of the FPT for integral threshold is always less than that for the standard threshold (also see Supplementary Fig.~S1). In addition, use of the integral thresholding scheme for timekeeping has the added benefit of being more robust, since there is less sensitivity to noise in the value of the threshold. To validate this, in Supplementary Fig.~S2 we show that while the variance of the FPT with standard threshold increases exponentially with the threshold value, the corresponding variance for the integral threshold increases linearly, i.e., much less dramatically.

An analytical solution for the FPT probability density of the time integral, $\mathcal{P(\t;\th)}$, is found as previously described. We provide a simple closed form solution for the FPT distribution of the integral (for large threshold values):
\begin{equation}
\begin{split}
	\mathcal P(\tau; \theta) &=  
%	- \mathcal L^{-1}\left\{  \frac{\partial_t\tilde G(\iota, t)}{\iota}\right\} = \mathcal L^{-1} \left\{ \frac{k_b\, e^{-\frac{\iota}{\iota+1}k_b \,t}}{\iota+1} \right\} = 
		k_b\, e^{-k_b\, \tau - k_d\,\theta}I_0\left(2\sqrt{k_b k_d\, \theta\,\tau}\right),
\end{split}
\end{equation}
where $I_0(x)$ is the modified Bessel function of the first kind of order 0~\cite{2014-gradshteyn-xq}. While the characteristic timescale of the FPT is set $1/k_{b}$, that for the integral threshold, $\th$, is set by $1/k_{d}$. Interestingly, if $\t$ is rescaled by $1/k_b$ then the shape of the distribution is determined by the single parameter, the rescaled value of the integral threshold, $k_{d} \,\theta$. Qualitatively, this distribution is unimodal, positively skewed, and has an exponential tail.

{\bf {\em Applications to pendulum-clocks.}} As a prototype for stochastic pendulum-clocks, we consider a model in which the stochastic timekeeper, $Q$, undergoes periodic oscillations on average (see Fig.~\ref{fig-osc}). We use \eq{birth_death}, with $k_{b}(t) = k_b [1+\epsilon \cos(\omega\, t)]$ and $0\leq \epsilon\leq 1$. As desired, the mean value oscillates periodically: $\la q(t)\ra =  k_{b}/k_{d} +  \epsilon\, k_{b} [k_{d} \cos (\omega t) + \omega \sin (\omega t)] /(k_{d}^2+\omega^2)$. We note that $\la q(t)\ra$ is phase-shifted with respect to $k_{b}(t)$, and the magnitude of the phase shift is frequency dependent.

A rich variety of behaviors is obtained for the FPT distribution of the time integral of $Q$ for stochastic pendulum-clocks. The specifics depend on the hierarchy of relevant timescales for a given set of parameters. First, in the limit of fast oscillations, i.e., for $\omega \gg k_{d}$, the effect of oscillations is washed out, and one recovers the results previously obtained for the hourglass timer, i.e., for $\epsilon = 0$. If the threshold phase corresponds to multiples of the oscillation time-period (a relevant scenario for systems with biological rhythms~\cite{2001-rensing-ff, 2017-johnson-zv}), the integral threshold provides excellent temporal specificity (Fig.~\ref{fig-osc}D). As shown in the hourglass timer case, the FPT distribution for the integral thresholding scheme has a coefficient of variation which is much smaller than the corresponding FPT distribution for the standard threshold (when both mean FPTs are kept equal).

%%%%%%%%%%%%%%%%%%%%%%begin%%%%%%%%%%%%%%%%%%%%%%
%%%%%%%%%%%%%%%%%%%%%%%end%%%%%%%%%%%%%%%%%%%%%%

Remarkably, a population of identical cells can spontaneously bifurcate into subpopulations of early and late responders, depending on hierarchy of timescales in the dynamics. In other words, the FPT distribution of the time-integral of $Q$ can become bimodal (or multimodal) for some choices of parameters (Fig.~\ref{fig-osc}C). This feature may be used by biological systems where it is beneficial to have biphasic response to a given input signal. This is surprising, since the result for the hourglass timer is always unimodal, the integral is a monotonically increasing function of time, and the probability density, $\rho(\lambda, t)$, is unimodal.

Physically, multimodality arises due to a bottleneck effect caused by small number fluctuations, in models with non-monotonic copy number dynamics for the biochemical timekeeper (e.g., the stochastic pendulum clock), if fluctuations relative mean are large enough to drive its numbers to zero in some realizations. The integral is constant or increasing slowly for these trajectories near the minimum of $\la q(t) \ra$ (Fig.~\ref{fig-osc}). The next reaction is likely to be a birth event, and the integrals will increase more rapidly once the minimum is cleared, thus leading to a bifurcation of the population into early and late responders. 

In general, increasing the integral threshold, the amplitude $\epsilon$, or the frequency $\omega$ (for small values compared to $k_d$) moves the distribution to the right and increases the distribution's width, the number of peaks, and their amplitudes. The width and modality of the distribution decrease as the threshold approaches an integer multiple of the integral of the mean over one period, while these increase as it departs from such values (Fig.~\ref{fig-osc}C and D).

{\bf {\em Biological realization and discussion.}} We identify a mechanism that allows a cell to keep track of the time integral of the copy numbers of relevant biochemicals. We saw in the model in \eq{birth} that a stochastic variable $R$, whose birth-rate is proportion to $q$, provides a natural representation for the time integral of $q(t)$. In fact, for large $k_r$, $r$ has the same distributions as the integral of $q$~\cite{2005-karlis-fk}. (Since the Poisson parameter, $\lambda$, is equal to $k_r \int q(t) dt$, for large $k_r$, the corresponding Poisson distribution approaches a delta function, and the distribution $P(r,t)$ approaches $\rho(r,t)$). 

Cells may take advantage of the integral thresholding scheme with regulatory networks in which a relatively shortly lived timekeeping biochemical, $Q$, increases the propensity of production of another biochemical, $R$, which is much more stable (long lived). We expect such pairs of biochemicals (integrands and integrals) to be ubiquitous in gene regulatory networks~\cite{2015-milo-dp}.

A  commonly occurring example of such a pair of biochemicals is a messenger RNA and its corresponding protein. The lifetime of a messenger RNA in bacterial cells ranges from a fraction of a minute to half an hour, while the lifetime of corresponding proteins typically exceeds the generation time of the bacteria in growth phase (order of tens of minutes)\cite{2015-milo-dp}. Thus, proteins copy numbers are effectively time-integrals of the corresponding messenger RNA. Given a specific model of stochastic gene expression, the FPT distribution for the average copy number of the proteins can be computed using the framework provided here.

It is has been directly observed that individual cell sizes inform when cells divide~\cite{2014-iyer-biswas-vo} . Moreover, though cell size growth is stochastic, it has been observed cell sizes increase strictly monotonically~\cite{2014-iyer-biswas-vo, 2014-iyer-biswas-lt}. Thus, cell size is a candidate for a time-integrated cellular variable, which is thresholded (by an adder, timer or sizer scheme)~\cite{2014-iyer-biswas-vo, 2016-iyer-biswas-pz, 2014-amir-xe}, to control the timing of cell division. 

As previously remarked, the experimental challenges in obtaining high quality datasets for time-courses of individual cell dynamics have resulted in lacunae in our understanding of stochasticity in the {\em timing} of key cellular events~\cite{2016-iyer-biswas-pz}. However, given recent developments in single-cell technologies~\cite{2013-norman-aw, 2014-iyer-biswas-vo, 2015-lambert-sb, 2016-potvin-trottier-sv, 2014-melendez-so, 2014-bisaria-du}, we anticipate that experimental validation of the integral thresholding scheme proposed here will be forthcoming. 

\begin{acknowledgments}
We thank Rudro Biswas and Sid Redner for insightful discussions. S.I-B. thanks the Santa Fe Institute, where a portion of the work was completed, for hospitality. We acknowledge financial support from Purdue University Startup Funds and the Purdue Research Foundation. S.I-B. thanks the W. M. Keck Foundation for financial support during early stages of this work. 
\end{acknowledgments}

\section*{Author Contributions}
SI-B conceived of and designed research, and developed the theoretical framework. FJ, MV and SI-B performed calculations and simulations, and wrote the paper.

\onecolumngrid
\clearpage

%\vspace{1cm}
\begin{center}
{\bf\Large Supplemental Information}
\end{center}
%\vspace{0.1cm}
\setcounter{secnumdepth}{3}  
\setcounter{section}{0}
\setcounter{equation}{0}
\setcounter{figure}{0}
\renewcommand{\theequation}{S-\arabic{equation}}
\renewcommand{\thefigure}{S\arabic{figure}}
\renewcommand\figurename{Supplementary Figure}
\renewcommand\tablename{Supplementary Table}
\newcommand\Scite[1]{[S\citealp{#1}]}
\makeatletter \renewcommand\@biblabel[1]{[S#1]} \makeatother

%%%%%%%%%%%%%%%%%%%%%%begin%%%%%%%%%%%%%%%%%%%%%%
\begin{figure}[h]
	\centering
	\includegraphics[width=.85\textwidth]{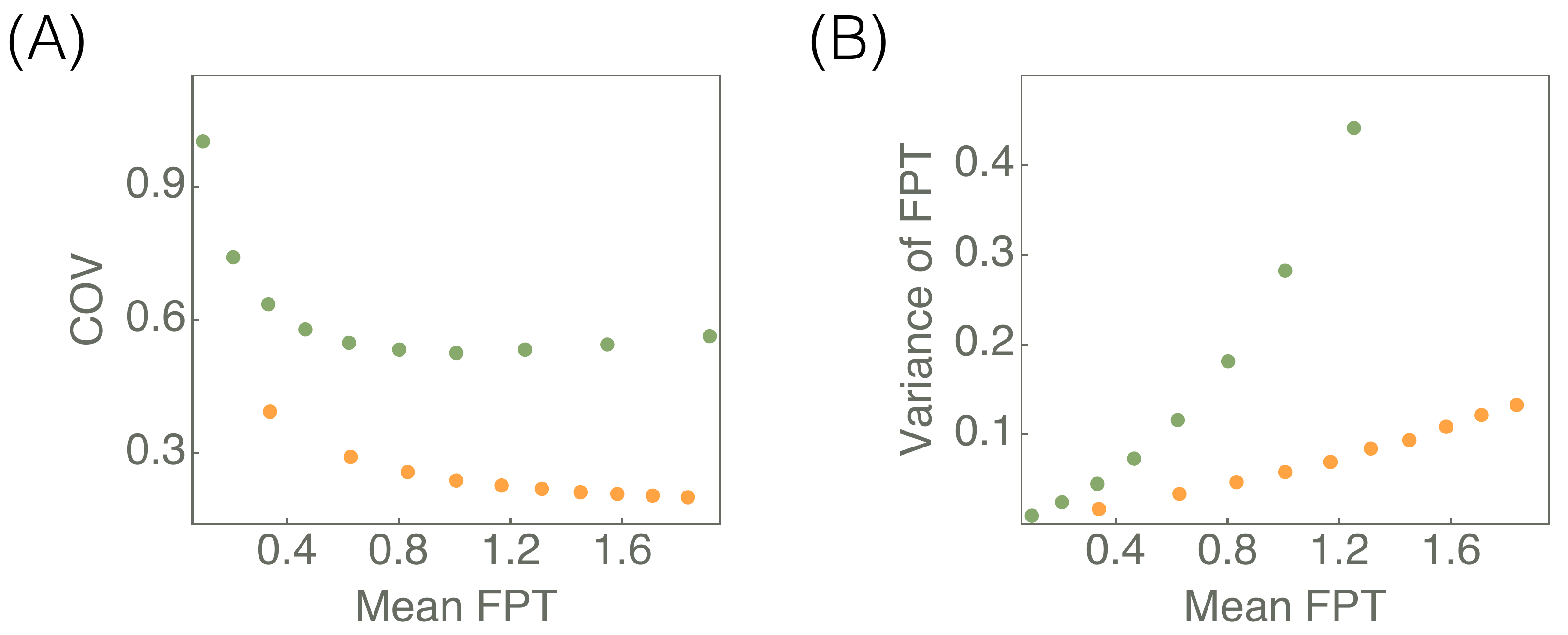}
	\caption{For the hourglass model, improvement in specificity is achieved by thresholding the time-integral instead of the original stochastic signal. (A) The coefficient of variation (COV) of the FPT of a birth-death process with constant birth rate (green markers), as a function of mean FPT, is contrasted with the trend for the COV of the FPT distribution of its time-integral (orange markers).  The COV of FPT of the integral decays monotonically and is always less than that for the original signal. (B) The variance of the FPT of the birth-death process grows exponentially with its mean (green markers) and is significantly larger than the variance of the FPT of the integral of the signal (orange markers). Parameters used: $k_d = 1$, $k_b = 10$. The distribution of the FPT of the birth-death process is calculated by imposing absorbing boundary condition on the value of the threshold and calculating the probability flux through the absorbing boundary. The distribution of the FPT of the integral is calculated using the numerical Laplace inverse of the expression provided in the text.}
\end{figure}
%%%%%%%%%%%%%%%%%%%%%%%end%%%%%%%%%%%%%%%%%%%%%%

%%%%%%%%%%%%%%%%%%%%%%begin%%%%%%%%%%%%%%%%%%%%%%
\begin{figure}[h!]
	\centering
	\includegraphics[width=.88\textwidth]{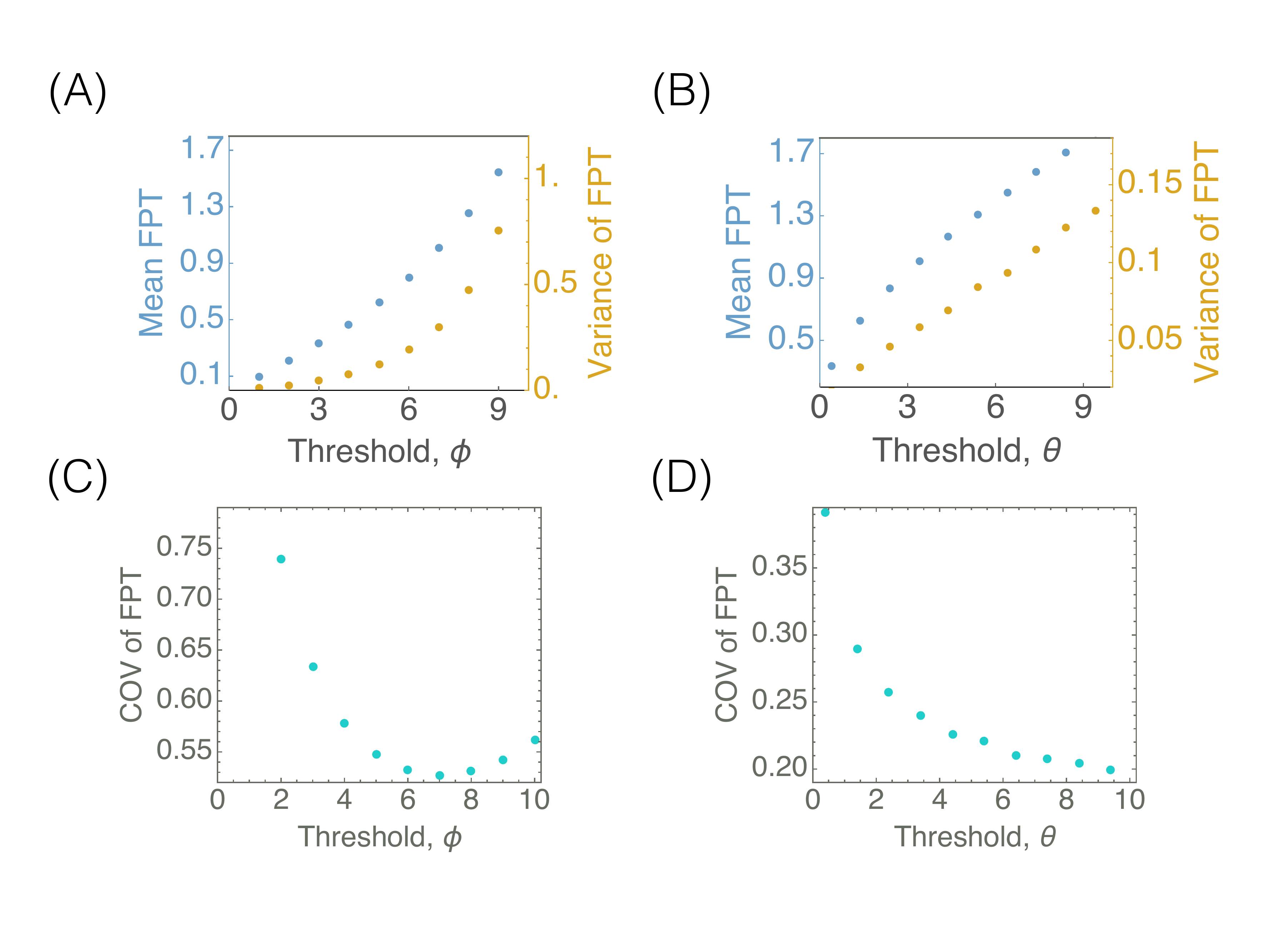}
	\caption{Robustness in implementation of the thresholding scheme is improved if the time-integral of a stochastic signal is thresholded instead of the original signal. (A) The mean and the variance of the FPT of a birth-death process both grow exponentially as a function of the threshold, $\phi$. The extreme sensitivity to the threshold value makes the control mechanism unstable with respect to perturbations to the system which affect the threshold value. (B) Both the mean and the variance of the FPT of the integral of the signal grow linearly with the threshold value, $\theta$, which allow the system to control the timing of the triggered event in a stable manner. (C) The COV of the FPT of a birth-death process changes non-monotonically with the value of its threshold, while (D) the COV of the FPT of the integral decays monotonically with the value of its threshold. Parameters used: $k_d = 1$, $k_b = 10$.}
\end{figure}
%%%%%%%%%%%%%%%%%%%%%%%end%%%%%%%%%%%%%%%%%%%%%%

%\bibliography{int_thresh}
\end{document}